\documentclass{article}

\usepackage[utf8]{inputenc}
\usepackage{graphicx}                       

\usepackage{subfigure}                      

\newtheorem{definition}{Definition}

\usepackage{xcolor}
\definecolor{dkgreen}{rgb}{0,0.6,0}
\definecolor{darkgreen}{rgb}{0,0.4,0}
\definecolor{gray}{rgb}{0.5,0.5,0.5}
\definecolor{mauve}{rgb}{0.58,0,0.82}
\definecolor{orange}{rgb}{1,0.49,0.24}

\usepackage{xcolor}

\usepackage{listings}
\usepackage[raggedright]{titlesec}

\usepackage{hyperref}  

\begin{document}

\title{Harnessing Historical Corrections to build Test Collections for Named Entity Disambiguation\thanks{This is a preprint of a paper accepted at TPDL 2018. The TPDL paper is available at: \url{https://doi.org/10.1007/978-3-030-00066-0_4}}}

\author{Florian Reitz \\ Schloss Dagstuhl LZI, dblp group, Wadern, Germany \\ florian.reitz@dagstuhl.de}
\providecommand{\keywords}[1]{\textbf{Keywords --} #1}

\date{}
\maketitle

\begin{abstract}
Matching mentions of persons to the actual persons (the name disambiguation problem) is central for several digital library applications.
Scientists have been working on algorithms to create this matching for decades without finding a universal solution.
One problem is that test collections for this problem are often small and specific to a certain collection.
In this work, we present an approach that can create large test collections from historical metadata with minimal extra cost. 
We apply this approach to the DBLP collection to generate two freely available test collections. 
One collection focuses on the properties of defects and one on the evaluation of disambiguation algorithms. 
\end{abstract}


\section{Introduction}\label{sec:Introduction}

Digital libraries store list of names which refer to real world persons (e.g., the authors of a document). 
There are many applications that need a map between these names and the real world persons they refer to.
E.g., a project might want to create author profiles which list all publications of a person.
These profiles are useful for users who are looking for works of a specific person.
They are also the basis of many approaches that try to measure the performance of researchers and institutions. 
Mapping author mentions and persons is difficult.
The name itself is not well-suited to refer to a person and many metadata records provide limited additional information such as email and institution name.
Therefore, many profiles list publications from different persons (a homonym defect) or publications of one person are listed in different profiles (a synonym defect).
Correct author disambiguation in bibliographic data has been the subject of intensive research for decades.
For an overview on algorithmic approaches, see the survey by Ferreira et al. \cite{ferreira2012brief}.
For manual strategies, see the paper by Elliot\cite{elliot2010survey}.
New approaches are proposed every year. 

Many approaches concentrate on reclustering the existing data.
I.e., the algorithm is provided with all mentions of persons and clusters these mentions into profiles. 
An advantage of this approach is that potentially wrong profiles can be ignored.
The problem is that reclustering ignores disambiguation work which has already been invested into the collection.
For a living collection, we can expect a significant amount of manual and automatic work invested in the correctness of data.
With an increasing number of open data projects, we can expect more user participation in the disambiguation process (e.g., users use ORCID to manage a clean publication profile).
We will also see larger collections as sharing and incorporating data will become easier. 
For a large and volatile collection, reclustering might be algorithmically unfeasible.
An alternative disambiguation task is to identify profiles which are likely defective.
These profiles can then be corrected automatically or checked by staff or in a crowd-sourcing framework. 
One problem of developing algorithms for this task is the lack of suitable test collections for evaluation.
Traditional test collections, such as as the set provided by Han et al.\cite{HanZG05} consist of mentions with the same name without any known author profiles.
This is not useful for the defect detection task.
In addition, there are the following problems:
(1) Classic test collections are small. See Müller et al.\cite{MuellerRRx} for an overview.
The largest collections discussed have several 10 thousand mentions, while collections like dblp list more than 10 million mentions.
This makes classic collections unusable for runtime evaluation.
(2) Classic test collections cannot be used to study properties of defects.
A study of properties can reveal new approaches to match mentions and persons.
It can also show differences between collections which need to be considered.
Known defects can also be used as training set, e.g., for deep learning applications.
In this work, we describe two alternative test collections which try to overcome these problems.

Creating a classic test collection is expensive, mainly because it requires manual cleaning of author profiles. 
However, for a large digital library, we can expect that a number of defects have already been corrected.
We extract these corrections from historical data and use them as examples of defective data. 
Since the defects have been corrected, we also know a (partial) solution to the defect.
Based on the defects, we build two test collections.
Our goal is to provide as much contextual information for each defect as possible.
One of the collections focuses on individual defects.
The other test collection focuses on the defect detection task in a large collection itself.
Harnessing historical corrections has several advantages:
(1) The collections we obtain are large compared to traditional test collections and created with minimal additional cost. 
(2) Unlike classic test collections, our approach is well-suited to study the properties of defects.
This can lead to a better understanding of quality problems and can be used when designing new disambiguation algorithms.
As defect corrections can be triggered in may different ways, we obtain a large variety of defects. 
(3) The framework we present can be used for all digital libraries that provide historical metadata. 
This might provide us with specific test collections which can be used to adjust algorithms to the properties of individual collections.
The main contributions of this work are:
\begin{itemize}
 \item We present a framework to create test collections for the defect detection task from historical data.
 \item We use the framework to create an open test collection based on the dblp collection.
\end{itemize}

After discussing related work, we describe how to build test collections from historical defect corrections (Section \ref{sec:framework}).
In Section \ref{sec:dblp}, we apply the framework to dblp and discuss possible applications of the test collections.

\section{Related Work}\label{sec:relwork}

Building test collections for the name disambiguation task is difficult. 
The usual approach is to select a small potion of data from a collection and clean it thoroughly.
This requires manual work which leads to small test collections. 
E.g., the often used test collection by Han\cite{HanZG05} consists of about 8400 mentions (which roughly equals publications), the KISTI collection by Kang et al.\cite{KangKLJY11} consists of about 32000 mentions.
For comparison, the dblp collection listed about 10 million mentions in March 2018.
For an overview, see the work of Müller et al.\cite{MuellerRRx}.
Most test collections consist of two sets, the challenge which is presented to the algorithms and the solution which contains the correct clustering of mentions into profiles.
Algorithms are judged by how close they can approximate the solution.
E.g., for Han et al., the challenge consists of publications from authors with common names (such as \textit{C. Chen}).
The authors first name is abbreviated to increase the difficulty. 
Data that could not be manually disambiguated was discarded.
Most test collections provide the basic bibliographic metadata such as title, name of coauthors and publication year.
This creates compact test collections but provides very little context information.
E.g., the collections only contain partial information on the coauthors because most of their publications are not part of the test collection.
This might be unfair to algorithms that use the coauthor network to disambiguate authors.

Since manual disambiguation for test collections is expensive, there have been several attempts to harness work which has  already been invested into a collection.
Reuther\cite{Reuther2007} compared two states of dblp from different years to see if publications had been reassigned between author profiles.
Reuther gathered the publications of these profiles for a test collection that focuses on corrections of synonym defects.
Momeni and Mayr\cite{MomeniM16} build a test collection based on homonym profiles in dblp.
When dblp notices that a name is used by several authors, the author mention is appended by a number. 
E.g., \textit{Wei Wang 0001}, ..., \textit{Wei Wang 0135}. 
Momeni and Mayr build a challenge by removing the suffixes.
The full name (including suffix) is used as solution.
Müller et al.\cite{MuellerRRx} describe how a test collection can be built by comparing the manual disambiguation work of different projects. 
For all these approaches, the data presentation is record based like for the classic test collections.

\section{Extracting Test Collections from Historical Metadata}\label{sec:framework}

Assume that publications for a person \textit{John Doe} are listed in author profiles \textit{J. Doe} and \textit{John Doe}. 
If that defect is uncovered, it will most likely be corrected.
Many collections attempt author disambiguation when data is added.
In the example, this did not work which indicates that this defect is not trivial which makes it interesting for defect detection algorithms.
We use the state of the collection before the correction as challenge to see if an algorithm can detect the presence of a problem and possibly propose a solution.
Using historical corrections has a number of advantages:
(1) The corrections can come from a number of different sources which can include defects with different properties.
E.g., for dblp, a significant amount of defects are reported by uses\cite{Reitz011}.
(2) At the moment of the correction data was available that might not be available today.
In 2014 Shin et al. \cite{ShinKCK14} reconsidered the data of Han et al. from 2005\cite{HanZG05} and determined that more than 22\% of their gold standard could not be confirmed.
In 2005 that verification was probably possible (web pages went off line, publishers become defunct ...). 
We will now describe an extraction approach for historical defects and how to build test collections on top of them.
In Section, \ref{sec:dblp} we show how this approach can be applied to the dblp bibliography to create test collections.

\subsection{Identifying Corrections in Historical Data}\label{sec:corrections}

\begin{figure}
\begin{center}
\includegraphics[width=75mm]{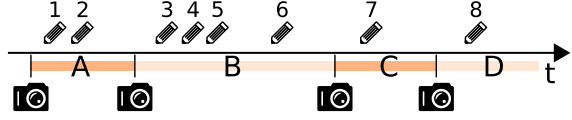}
\end{center}
\caption{Observer-based framework for historical metadata. \label{img:observer}}
\end{figure}

For our approach, we need suitable historical metadata.
Locating this data turned out to be difficult.
If a project provides historical data, it is often necessary need to use secondary data sources like backup files. 
For these data, we cannot expect to capture every correction separately. 
Instead, we obtain observations of the data.
The points of observation depend on the underlying data, e.g., the times the backups were created.
If observations are far apart and edits frequent, we might not be able to extract individual corrections. 
Figure \ref{img:observer} shows an example with edits and observations. 
In this case, we obtain four states, $A, \dots, D$ even tough there are more edits.
E.g., edits 3,4 and 5 might be merged into one observed correction. 
For each dataset, we need to determine if the observation allows for reasonable correction analysis.
E.g., for dblp, we use a collection\cite{hoffmann_oliver_2018_1213051} that has nightly observations. 
This granularity allowed for reliable correction extraction. 
For the Internet Movie Database (IMDB), we obtained weekly observations\footnote{\url{ftp://ftp.fu-berlin.de/pub/misc/movies/database/frozendata}}.
This made interpreting the data difficult.

If there is sufficient historical data, we can extract corrections.
Most digital libraries provide interpretations of their person mentions.
We call this interpretation a \textbf{profile}.
A profile contains the mentions (publications) that the digital library thinks are created by the person represented by the profile.
The interpretation can be based on the name directly (as in dblp) or based on an identifier assigned to the mention.
We require that the interpretation is contained in the historical metadata.
I.e., we can reconstruct historical profiles.
As explained above, some profiles will be defective (i.e., they deviate from the real person's work list).
Let $t_1 < t_2$ two time points of observation and let $p\langle t\rangle$ be the set of mentions that is assigned to profile $p$ at time $t$.
We can observe two types of relations between profiles from different observations:

\begin{definition}
\label{def:refSucPre}
Let $p_1$, $p_2$ be two profiles. We call $p_1$ \textbf{reference predecessor} of $p_2$ if $\exists \; m \in p_1\langle t_1 \rangle : m \in p_2\langle t_2 \rangle$. We call $p_2$ 
We call $p_1$ \textbf{consistent predecessor} of $p_2$ if $ p_1\langle t_1 \rangle \subseteq p_2\langle {t_2} \rangle$.
\end{definition}

There are two candidates for a defect correction:

\begin{enumerate}
 \item A profile $p$ has two or more reference predecessors.
 \item A profile $p$ has two or more reference successors.
\end{enumerate}

In case (1), $p$ was represented by multiple profiles before.
If we assume that $p$ is correct now, the successors where synonyms. 
Similarly, in case (2) we observe the correction of a homonym defect.

We can categorize modifications to profiles as follows:

\begin{definition}
\label{def:mergegroup}
Let $p$ be a profile and $t_1 < t_2$ two time points of observation.
Let $P := p_1, \dots, p_k$ be the reference predecessors of $p$ with respect to $t_1$ and $t_2$.
We call $P$ a \textbf{merge group} if $k > 1$ and $\forall 1 \leq s \leq k: p_s\langle {t_2} \rangle = \emptyset \lor p_s = p$.
\end{definition}

Between time $t_1$ and $t_2$, mentions in $p_1, \dots, p_k$ were reassigned to $p$.
These profiles, except $p$ itself, do not have any mention left.
We can consider a similar correction for homonyms:

\begin{definition}
With $p$, $t_1$ and $t_2$ as above.
Let $P := p_1, \dots, p_k$ be the reference successors of $p$ with respect to $t_1$ and $t_2$.
We call $P$ a \textbf{split group} if $k > 1$ and $\forall 1 \leq s \leq k: p_s\langle {t_1} \rangle = \emptyset \lor p_s = p$.
\end{definition}

For a merge, we demand that the merged profiles are no longer referenced in the library.
Similarly, we demand for a split that \textit{new} profiles do not contain a mention at $t_1$.
In addition to that, we need to consider a combination of merge and split, a distribute.
In this case, a mention is moved from one profile to another without creating an empty profile.
Distributes are different from merges and splits as both profiles are represented before and after the correction.
An algorithm which aims to correct this defect must determine which mentions are to be reassigned.
If both profiles exist before the correction, the algorithm can use their properties to determine if a mention needs reassignment.
This might be easier than detecting completely merged mentions. 

Merge and split groups can combine multiple corrections. 
This can create artifacts with the observation framework.
Assume that there are two merge corrections ${p_1, p_2} \rightarrow {p_1}$ and ${p_1, p_3} \rightarrow {p_1}$.
If these operations occur between two observations, we obtain a merge group ${p_1, p_3} \rightarrow {p_1}$.
If the observation occurs between the two corrections, we obtain two merge groups. 
To avoid splitting groups that are related, we group merge and split groups if they occur in direct succession and have at least a common profile. 

\subsection{Structure of Test Collections}\label{sec:teststructure}

The extracted corrections can now be transformed into test collections. 
Classic test collections list the records of the disambiguated authors.
This creates compact collections however it provides very little context information.
For our test collections, we have the following goals:

\begin{enumerate}
 \item The collection should allow to study the properties of defects. In particular, they should provide the context information which is commonly used by disambiguation algorithms such as parts of the coauthor network.
 \item The collection should facilitate the development of algorithms that search for defects in an existing name reference interpretation. As opposed to collections that aim at algorithms that completely recluster author mentions.
 \item The collection should support a runtime-based evaluation. 
 \item The collection should be of manageable size.
\end{enumerate}

To meet these goals, we create two collections that both differ from classic collections:
A case-based collection that lists the individual corrections as small graphs.
An embedded collection that integrates the detected defects into the total collection.
We now discuss the general structure of the two collections. 

\subsubsection{Case-based Collection}
\begin{figure}
\begin{center}
\includegraphics[width=70mm]{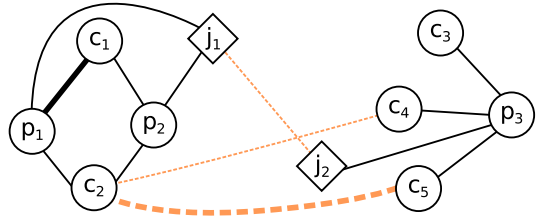}
\end{center}
\caption{Relations between metadata entities as a graph. \label{img:relation}}
\end{figure}

The case-based collection consists of isolated test cases that are directly derived from the observed corrections.
For each correction, we provide two files.
One file contains the state of the digital library directly before the correction, the other file contains the state right after the correction.
The primary purpose of the case-based collection is to study the properties of defects.
This requires that a certain context is provided.
E.g., many disambiguation algorithms use common coauthors as evidence that there is a relation between two mentions.
Classic test collections provide this information but they give no information about the relations between the coauthors.
Consider Figure \ref{img:relation} with synonymous profiles $p_1, p_2, p_3$, coauthors $c_1, \dots, c_5$ and journals $j_1$ and $j_2$.
$p_1$ and $p_2$ are strongly related by two common coauthors and a journal.
$p_3$ is not in a direct relation to $p_1$ and $p_2$.
The black solid lines represent the data available from a classic test collection. 
The dashed lines represent contextual data that is not in the test collection.
$c_2$ and $c_5$ collaborated. 
However, that relation is defined by publications outside the test collection.
Studying these indirect relations might help to develop a better disambiguation algorithm.
 
Obviously, we cannot provide the complete metadata context (e.g., the complete coauthor graph) for each test case.
To provide at least local context, we code the test case as a graph.
Consider again Figure \ref{img:relation}.
Assume that $p_1, \dots, p_3$ profiles are part of a merge group. 
We create a graph as follows:
We add nodes for all corrected profiles ($p_1, \dots, p_3$).
We call these nodes primary nodes.
We add a node for each entity that is in relation to a primary node (e.g., the coauthors).
The set of available entities depends on the underlying digital library.
Other entities might be conferences/journals or common topics.
We then add an edge for each known relation between these nodes. 
The context is provided by the edges between the nodes that are no primary nodes.
The types of relations depend on the data in the digital library.
The edges can be weighted which makes it possible to convey the strength of a relation without providing lots of information.  

We encode the graphs in XML. 
For each edge and node, we can provide properties in (key, value) form. 
The following example shows a document-type node that has title and publication year information.
A similar notation is used for edges.

\begin{lstlisting}
<node label="DOCUMENT" id="doc1">
     <property key="year" value="1999"/>
     <property key="title" value="The Ultrasonic Navigating."/>
</node>
\end{lstlisting}

\subsubsection{Embedded Collection}

The defects of the case-based collection are too small to pose a runtime challenge.
They also do not provide a full context.
Some disambiguation algorithms require a full coauthor graph\cite{jcst/SunSKNY16}\cite{jdiq/FanWPZL11} which is not available from the local context of the individual cases. 
The embedded collection solved these problems.
It consists of two components: 
(1) A full copy of the collection's metadata at a certain point, provided as metadata records.
(2) An annotation of detected defects in this version which are corrected later.
Algorithms need to process the full collection but have also access to all data.

Since we provide the full version of the metadata, it is not possible to use a dense observation framework.
I.e., for detecting defect corrections, we need to compare states of the collection which are some time apart (e.g., a full year).
This will create a sufficient number of defects to be annotated.
However, the long periods between the states of the data set makes overlapping corrections more likely.
Assume that we observe a distribute operation between author profiles $p_1$ and $p_2$ (publications are moved between these profiles).
Further assume that profile $p_3$ is merged into profile $p_1$.
For a dense observation framework, there are many different ways in which these operations can be performed.
In a slightly different situation, we might have observed a distribute between $p_2$ and $p_3$ and a merge of $p_1$ into $p_3$.
For a sparse observation framework, these corrections will most likely be merged together. 
This does not affect the presence of a defect (which should be detected by the algorithm) but makes the embedded collection unsuited to analyze individual corrections.
The metadata of the collection can be provided in any way, e.g., as metadata records.
Unlike the case-based collection, this might make importing the data easier for some approaches.
The annotations are provided as simple XML-Files. 
The example below shows a small split case. 
\texttt{doc1}, \texttt{doc2}, \texttt{p1} and \texttt{p2} are identifiers taken from the underlying collection.

\begin{lstlisting}
<source>
   <profile authorid="p1">
      <signature pkey="doc1" pos="1" surface="B. Doe"/>
      <signature pkey="doc2" pos="0" surface="B. Doe"/>
   </profile>
</source>
   <target>
      <profile authorid="p1">
         <signature pkey="doc1" pos="1" surface="Bob A. Doe"/>
      </profile>
      <profile authorid="p2">
         <signature pkey="doc2"  pos="0" surface="Bob B. Doe"/>
   </profile>
</target>
\end{lstlisting}

\subsection{Biases and Limitations}\label{sec:bias}

The test collections we present here are different from the classic test collections as they do not provide a full gold standard.
This means:
(1) They provide examples of errors but have no examples of guaranteed correct data which could be used to detect false positives.
(2) The corrections might be partial.
See below for an example.
In Section \ref{sec:dblp}, we will very briefly discuss scenarios in which the collections can be used. 
It is important to note that these collections will not replace classic test collections but complement them. 
E.g., to study defects or to evaluate runtime.
Apart from the evaluation method itself, our approach has intrinsic biases which cannot be fully mitigated.
In this section, we will discuss the most relevant points.
Each of these threats to validity must be considered before undertaking a study based on historical defect corrections. 

\textbf{Assumption:} Corrections improve data quality.
We assume that a correction replaces defective data values with correct values. 
Obviously, there is no guarantee for that as the changes related to the correction can also introduce errors. 
The likelihood of introducing new errors depends on the data curation process of the individual projects.
Some projects use trained teams while others rely on direct or indirect user contribution.
On the other hand, user contribution might be vandalism.
In any case, we will obtain a number of partially or fully defective corrections.

\textbf{Assumption:} Corrections completely remove defects.
A correction might remove a defect only partially. 
Assume that one profile contains publications from author $A$, $B$ and $C$.
A correction (a split) might extract the publication of $A$ but leave the publications of $B$ and $C$ behind.
The original profile is still a homonym.
If we build a test collection based on partial corrections, we must allow for a case where an algorithm finds the whole correction.
This means that there is no gold standard solution to our test collection as there is for classical test collections.
We need to define specific evaluation metrics to handle this situation.
In case of a study of defect properties, we must also consider that some corrections are only partial. 

\textbf{Assumption:} The corrected defects are representative of the set of all defects.
Our approach is biased by the way defects are detected in the underlying data set.
Assume that a project applies a process which is good at finding defects with property $\mathcal{A}$ but can barely handle defects with property $\mathcal{B}$. 
In this case, defects with property $\mathcal{A}$ would be overrepresented and many defects with property $\mathcal{B}$ would be missing. 
It is also possible that a project is aware of a defect but does not fix it because it has a low priority. 
Again, it is unclear how community contribution can mitigate this problem. 
For all studies, we must assume that error classes exist that are significantly underrepresented.

\section{Test Collections based on dblp}\label{sec:dblp} 

We apply the framework described above to the dblp bibliography\footnote{\url{https://dblp.org}}. 
The collections are published under an open license\cite{reitz_florian_2018_1215650}.
The dblp project gathers metadata for publications in computer science and related fields.
In March 2018, there were 4.1 million publications and 2 million profiles.
Dblp creates nightly backups of its data which are combined into a historical data file\cite{hoffmann_oliver_2018_1213051}.
This file can be used to trace modifications to the metadata records between June 1999 and March 2018.

\subsection{Application of the Framework}

Dblp has two mechanisms to match author mentions with observed entities.
(1) The name itself. The name might be appended with a numeric suffix such as \textit{Wei Wang 0050}.
(2) Authority records which map names to person entities.
The authority records are part of the historical data.
I.e., we can track changes to the authority data as well.

We use three different types of entities for the graphs of the case-based collection:
\textit{Document}, \textit{Person} and \textit{Venue} (journal or conference series)
The primary function of \textit{Document} is to provide the standard metadata such as title and year of publication.
We model six different relations.
\textit{Created}/\textit{Contributed} (Person $\rightarrow$ Document, unweighted): The person is author/editor of that document/proceedings.
\textit{Co-Created}/\textit{Co-Contributed}  (Person $\leftrightarrow$ Person, weighted by number of common papers): The persons are authors/editors of at least one common paper.
\textit{Created-At}/\textit{Contributed-At} (Person $\rightarrow$ Venue, weighted by number of papers): The person is author of a paper that appeared at the venue / editor of a proceedings of the venue.
We decided to model editorship and authorship separately as they might have different implications for an algorithm.
Coauthorship usually implies cooperation while being coeditors (e.g., of a proceedings) can simply mean that the authors are active in the same field.
Weights are computed for the last date before the correction was observed. 
I.e., the weights represent the data which would have been available to an algorithm at that time.
We provide all properties for the documents that are listed in dblp.
However, we use the most recent data instead of the data available at the point of correction.
The main reason is to provide current weblinks to the publication pages of publishers.
Today, these links are mostly resolved via DOI. 
An algorithm can use the links to get additional information from the web.
The case-based collection contains 138,532 merges, 16,532 splits and 55,362 distributes.

For the embedded test collection, we considered the state of dblp at the beginning of a year. 
Table \ref{tab:embeddedSizes} lists the number of corrections for some combinations of different dates.
We do not consider states of dblp from before 2013 as the collection was small at that time and the number of possible corrections is negligible.
The number of corrections is small compared to the case-based collection.
The primary reasons are (1) short-lived defects that were introduced to the collection and corrected between the observations are missing.
(2) As discussed above, we might merge multiple corrections into one. 

\begin{table}
\caption{Number of identified corrections for different observation frameworks. \label{tab:embeddedSizes}}
\begin{center}
\begin{tabular}{r|r|r|r|r}
\textbf{observation dates}	& \textbf{split}	& \textbf{merge}	& \textbf{distribute}	& \textbf{all} \\
\hline
2013, 2017	& 2.207 	& 19,175 & 5,346	& 26,728\\
2015, 2017	& 1.536 	& 13,393 & 3,968	& 18,897\\
2017, 2018	&   978 	&  8,608 & 2,666	& 12,252\\ 
\end{tabular}
\end{center}
\end{table}

\subsection{Possible Applications}\label{sec:dblpappl}

As stated above, both test collections do not provide full solutions of the name disambiguation task.
Therefore, classic evaluations such as cluster alignment cannot be used to evaluate the approaches.
However, the embedded collection can be used to test runtime performance and the general ability to detect known defects in a collection. 
A simple evaluation strategy would be:
(1) Use classic test collections to obtain precision/recall/cluster alignment in a fully solved scenario.
(2) Check if the algorithm can handle the size of the embedded test collection.
(3) Measure how many defects in the embedded collection are detected.
This will filter out slow approaches and provide insides if the qualitative performance from the classic test collection translates to the embedded collection.

The case-based collection can be used to study properties of defects.
As an example, we considered how suitable names are for simple blocking approaches.
Blocking is a preprocessing step which partitions the data set into manageable bins. 
The idea is that similar names are placed in the same bin\cite{HanZG05}\cite{jasis/LevinKBJ12}.
Some approaches use a similar idea to compute similarity between mentions\cite{SantanaGLF15}.
Blocking is mostly recall-based so we can use the case-based collection to measure the performance.
We consider two variations of name-based blocking:
(1) Only the last name part is considered. 
E.g., from \textit{John Doe} use \textit{Doe}.
(2) The last name and the initial of the first name is used. Middle names are ignored. From \textit{John A. Doe} we use \textit{J. Doe}.
We use both approaches with and without considering case. 
For merge and distribute cases, we compute how many name pairs are placed in the same block. 
Table~\ref{tab:namepair_abbrev} shows the result for dblp, compared to results from a test collection we built on IMDB.
The data for dblp are from an older version of the test collection which covers the period 1999-2015.

Both blocking approaches are designed with the name abbreviation problem in mind. 
The approaches perform well for dblp with hit rates between 76.51\% and 79.1\%. 
This is due to the large number of abbreviated names in academic publications. 
While a hit rate of 0.791 is far from optional -- of all name pairs 21\% do not end in the same block -- it might be acceptable.
However, the results for IMDB are much worse, indicating that blocking strategies that work well for one project are not suited for other libraries.

\begin{table}
\begin{center}
\caption{Comparison of abbreviation.based similarity. The table shows percentage of pairs which are considered similar. \label{tab:namepair_abbrev}}
\begin{tabular}{l|c|c|c|c}
  &	\multicolumn{2}{c|}{\textbf{consider case}} & \multicolumn{2}{c}{\textbf{ignore case}} \\
\textbf{project} & \textbf{initial + last} &  \textbf{last} & \textbf{initial + last} & \textbf{last} \\
\hline
DBLP merge+dist & 76.51\% & 78.56\% & 77.10\% & 79.10\% \\
IMDB merge+dist & 46.24\% & 56.64\% & 47.15\% & 57.57\% \\
\end{tabular}
\end{center}
\end{table}

\section{Conclusion}\label{sec:Conclusion}

In this work, we described how historical defect corrections can be extracted and processed into test collections for the name disambiguation task.
The collections do not permit classical evaluation but provide insights into the nature of defects and allow evaluation of aspects which have been difficult to test so far.
At the moment, it is still difficult to find usable historical data for most collections.
We hope that with an increasing number of open collections, this problem will be solved.
At that point, it will be possible to create individual test collections.
Using different collections will provide more stable algorithms that do not depend on properties of the underlying data. 

\section*{Acknowledgements}
The research in this paper is funded by the Leibniz Competition, grant no. LZI-SAW-2015-2.
The author thanks Oliver Hoffmann for providing the data on which the dblp test collection is built and Marcel R. Ackermann for helpful discussions and suggestions.

\bibliographystyle{abbrv}
\bibliography{datasetPaper}

\end{document}